\documentclass[10pt, conference, compsocconf]{IEEEtran}
\usepackage{cite}
\usepackage{graphicx}
\usepackage{subfigure}
\graphicspath{ {images/} }
\usepackage{url}
\usepackage[super]{nth}
\usepackage{array}
\usepackage{multirow}
\usepackage{graphicx}
\usepackage{mathptmx}
\usepackage{commath}
\usepackage{algorithm}
\usepackage{algpseudocode}
\usepackage{changepage}
\usepackage{amsmath}
\usepackage[utf8]{inputenc}
\usepackage{etoolbox}
\usepackage{xcolor}
\makeatletter
\patchcmd{\@makecaption}
  {\scshape}
  {}
  {}
  {}
\makeatother
\usepackage{graphicx}
\usepackage{float}

\usepackage{listings}
\usepackage{verbatim}

\def\ie{{\it i.e.}\hspace{0.1pc}}
\def\eg{{\it e.g.}\hspace{0.1pc}}
\def\etal{{\it et al.}\hspace{0.1pc}}
\def\etc{{\it etc.}\hspace{0.1pc}}

\begin{document}

\title{Verified Instruction-Level Energy Consumption Measurement for NVIDIA GPUs}

\author
{
\IEEEauthorblockN
{
Yehia Arafa\IEEEauthorrefmark{1}, Ammar ElWazir\IEEEauthorrefmark{1}, Abdelrahman Elkanishy\IEEEauthorrefmark{1},  Youssef Aly\IEEEauthorrefmark{4}, Ayatelrahman Elsayed\IEEEauthorrefmark{1},\\ Abdel-Hameed Badawy\IEEEauthorrefmark{1}\IEEEauthorrefmark{2}, Gopinath Chennupati\IEEEauthorrefmark{2}, Stephan Eidenbenz\IEEEauthorrefmark{2}, and Nandakishore Santhi\IEEEauthorrefmark{2}\\
\vspace{0.1pt}
}
\IEEEauthorblockA
{
\IEEEauthorrefmark{1} New Mexico State University, Las Cruces, NM, USA\\
\{yarafa, ammarwa, anasser, aynasser, badawy\}@nmsu.edu\\
\IEEEauthorrefmark{4} Arab Academy for Science, Technology \& Maritime Transport, Alexandria, Egypt\\
\IEEEauthorrefmark{2} Los Alamos National Laboratory, Los Alamos, NM, USA\\
\{gchennupati, eidenben, nsanthi\}@lanl.gov\\
}
}

\maketitle

\begin{abstract}

GPUs are prevalent in modern computing systems at all scales. They consume a significant fraction of the energy in these systems. However, vendors do not publish the actual cost of the power/energy overhead of their internal microarchitecture. In this paper, we accurately measure the energy consumption of various PTX instructions found in modern NVIDIA GPUs. We provide an exhaustive comparison of more than 40 instructions for four high-end NVIDIA GPUs from four different generations (\textit{Maxwell, Pascal, Volta, and Turing}). Furthermore, we show the effect of the CUDA compiler optimizations on the energy consumption of each instruction. We use three different software techniques to read the GPU on-chip power sensors, which use NVIDIA’s NVML API and provide an in-depth comparison between these techniques.
Additionally, we verified the software measurement techniques against a custom-designed hardware power measurement. The results show that \textit{Volta} GPUs have the best energy efficiency of all the other generations for the different categories of the instructions. This work should aid in understanding NVIDIA GPUs' microarchitecture. It should also make energy measurements of any GPU kernel both efficient and accurate.

\end{abstract}

\begin {IEEEkeywords}
GPU Power Usage, PTX, NVML, PAPI, Internal Power Sensors, External Power Meters
\end{IEEEkeywords}

\section{Introduction}
\label{sec:intro}

Applications that rely on graphics processor units (GPUs) have increased exponentially over the last decade. GPUs are now used in various fields, from accelerating scientific computing applications to performing fast searches in data-oriented applications. A typical GPU has multiple streaming multiprocessors (SMs). Each can be seen as standalone processors operating concurrently. These SMs are capable of running thousands of threads in parallel. Over the last decade, GPUs' microarchitecture has evolved to be very complicated. However, the increase in complexity means more processing power. Hence, the recent development of embedded/integrated GPUs and their application in edge/mobile computation have made power and energy consumption a primary metric for evaluating GPUs performance. Especially that researchers have shown that large power consumption has a significant effect on the reliability of the GPUs~\cite{GPGPU_reliability}. Hence, analyzing and predicting the power usage of the GPUs' hardware components remains an active area of research for many years.

Several monitoring systems (hardware \& software) have been proposed in the literature~\cite{NVML,powersensor2, powerinsight, powermon} to measure the total power usage of GPUs. However, measuring the energy consumption of the GPUs' internal hardware components is particularly challenging as the percentage of updates in the microarchitecture can be significant from one GPU generation/architecture to another. Moreover, GPU vendors never publish the data on the actual energy cost of their GPUs' microarchitecture. 

In this paper, we provide an accurate measurement of the energy consumption of almost all the instructions that can execute in modern NVIDIA GPUs. Since the optimizations provided by the CUDA (\textit{NVCC}) compiler~\cite{nvcc10.1} can affect the latency of each instruction~\cite{arafaInstLatency}. We show the effect of the CUDA compiler's high-level optimizations on the energy consumption of each instruction. We compute the instructions energy at the PTX~\cite{ptx} granularity, which is independent of the underlying hardware. Thus, the measurement methodology introduced has minimum overhead and is portable across different architectures/generations. 

To compute the energy consumption, we use three different software techniques based on the NVIDIA Management Library (NVML)~\cite{NVML}, which query the onboard sensors and read the power usage of the device. We implement two methods using the native NVML API, which we call \textit{Sampling Monitoring Approach (SMA)}, and  \textit{Multi-Threaded Synchronized Monitoring (MTSM)}. The third technique uses the newly released CUDA component in the PAPI v.5.7.1~\cite{papi-5.7} API. Furthermore, we designed a hardware system to measure the power usage of the GPUs in real-time. The hardware measurement is considered as the ground truth to verify the different software measurement techniques.

To the best of our knowledge, we are the first to provide a comprehensive comparison of the energy consumption of each PTX instruction in modern high-end NVIDIA GPGPUs. Furthermore, the compiler optimizations effect on the energy consumption of each instruction has not been explored before in the literature. Also, we are the first to provide an in-depth comparison between different NVML power monitoring software techniques. 

In summary, the followings are this paper \textbf{contributions}:

\begin{enumerate}
\item Accurate measurement of the energy consumption of almost all PTX instructions for four high-end NVIDIA GPUs from four different generations (\textit{Maxwell, Pascal, Volta, and Turing}).

\item Show the effect of CUDA compiler optimizations levels on the energy consumption of each instruction.

\item Utilize and Compare three different software techniques (\textit{SMA, MTSM, and PAPI}) to measure GPU kernels' energy consumption.

\item Verify the different software techniques against a custom in-house hardware power measurement on the \textit{Volta TITAN V} GPU.

\end{enumerate}

The results show that \textit{Volta TITAN V} GPU has the best energy efficiency among all the other generations for different categories of the instructions. Furthermore, our verification show that \textit{MTSM} leads to the best results since it integrates the power readings and captures the start and the end of the GPU kernel correctly.

The rest of this paper is organized as follows: Section~\ref{sec:background} provide a brief background on NVIDIA GPUs' internal architecture. Section~\ref{sec:methodology} describes the methodology of calculating the instructions energy consumption. Section~\ref{sec:software} depicts the differences between the software techniques. Section~\ref{sec:hardware} shows the in-house direct hardware design. In Section~\ref{sec:results} we present the results. Section~\ref{sec:related} shows the related work and finally, Section~\ref{sec:conc} concludes the paper.

\section{GPGPUs Architecture}
\label{sec:background}

GPUs consist of a large number of processors called \textit{Streaming Multiprocessor} (SMX) in CUDA terminology. These processors are mainly responsible for the computation part. They have several \textit{scalar cores}, which has some computational resources, including fully pipelined integer Arithmetic Units (ALUs) for performing 32-bit integer instruction, Floating-Point units (FPU32) for performing floating-point operations, and  Double-Precision Units (DPU) for 64-bit computations. Furthermore, it includes Special Function Units (SFU) that executes intrinsic instructions,  and Load and Store units (LD/ST) for calculations of source and destination memory addresses. In addition to the computational resources, each SMX is coupled with a certain number of warp schedulers, instruction dispatch units, instruction buffer(s), and texture and shared memory units. Each SMX has a private L1 memory, and they all share access to L2 cache memory. The exact number of SMXs on each GPU varies with the GPU's generation and the computational capabilities.

GPU applications typically consist of one or more kernels that can run on the device. All \textit{threads} from the same kernel are grouped into a \textit{grid}. The grid is made up of many \textit{blocks}; each is composed of groups of 32 threads called \textit{warps}. Grids and blocks represent a logical view of the thread hierarchy of a CUDA kernel. Warps execute instructions in a SIMD manner, meaning that all threads from the same warp execute the same instruction at any given time.


\section{Instructions Energy Consumption}
\label{sec:methodology}

We designed special micro-benchmarks to stress the GPU to be able to capture the power usage of each instruction.

We used Parallel-Thread Execution (PTX)~\cite{ptx} to write the micro-benchmarks. PTX is a virtual-assembly language used in NVIDIA’s CUDA programming environment. PTX provides an open-source machine-independent ISA. The PTX ISA itself does not run on the device but rather gets translated to another machine-dependent ISA named Source And Assembly (SASS). SASS is not open. NVIDIA does not allow writing native SASS instructions, unlike PTX, which provides a stable programming model for developers. There have been some research efforts~\cite{bare-metal:2017, assemblers_maxwell} to produce assembly tool-chains by reverse engineering and disassembling the SASS format to achieve better performance. Reading the SASS instructions can be done using CUDA binary utilities (\textit{cuobjdump})~\cite{binary_utilities}. The use of PTX helps control the exact sequence of instructions executing without any overhead. Since PTX is a machine-independent, the code is portable across different CUDA runtimes and GPUs.

\begin{figure}[tbp!]
      \centering
      \includegraphics[width=0.9\linewidth]{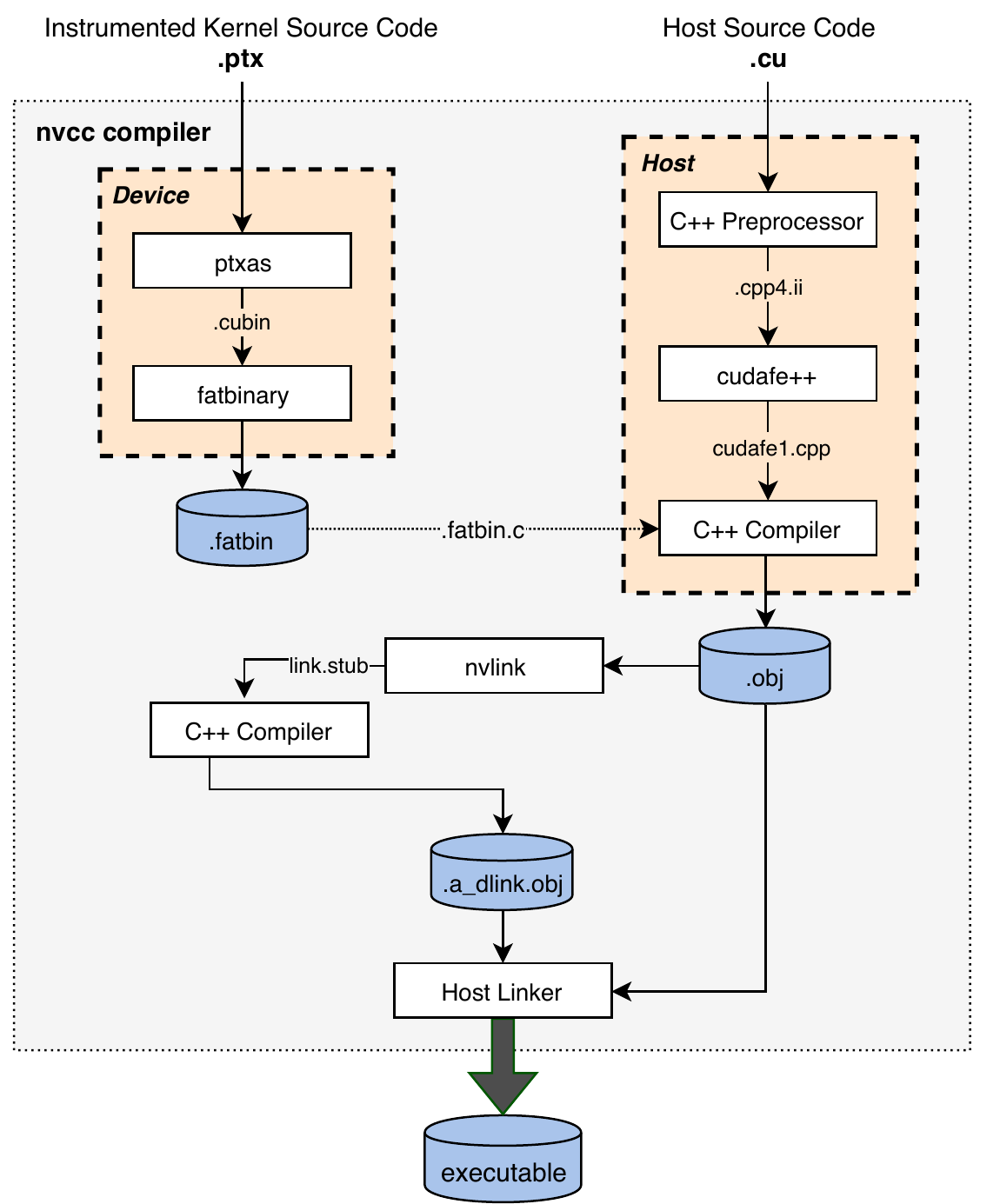}
      \caption{An Overview of the Compilation Procedure}
      \label{fig:workflow}
\end{figure}

Figure~\ref{fig:workflow} shows the compilation workflow, which leverages the compilation trajectory of the NVCC compiler. Since the PTX can only contain the code which gets executed on the device (\textit{GPU}), we pass the instrumented PTX device code to the NVCC compiler for linking at runtime with the host (\textit{CPU}) CUDA C/C++ code. PTX optimizing assembler (\textit{ptxas}) is first used to transform the instrumented machine-independent PTX code to a machine-dependent (\textit{SASS)} instructions then to a CUDA binary file (\textit{.cubin}). The binary is used to produce a \textit{fatbinary} file, which gets embedded in the host C/C++ code. An empty kernel gets initialized in the host code, which is then gets replaced by the instrumented PTX kernel, which has the same header and the same name inside the (\textit{.fatbin.c}). The kernel is executed with one block one thread.

\definecolor{backcolour}{rgb}{0.961,0.961,0.961}
\definecolor{code1}{rgb}{0.125,0.302,0.553}
\definecolor{code2}{rgb}{0.808,0.412,0}
\definecolor{codegray}{rgb}{0.5,0.5,0.5}

\lstdefinelanguage{PTX}
{
  morekeywords={ld, st, mov, add, sub, div, setp, bra},
  ndkeywords={global, param, ret, reg}
}

\begin{figure}[t!]
    \centering
           \begin{lstlisting}[language=PTX,
                      numbers=left, 
                      xleftmargin=0.26in, 
                      xrightmargin=0.14in, 
                      basicstyle=\scriptsize,
                      frame=single, 
                      backgroundcolor=\color{backcolour},
                      numberstyle=\scriptsize\color{codegray},
                      keywordstyle=\color{code1},
                      ndkeywordstyle=\color{code2}
                     ]
 .visible .entry Div(.param .u64 Div_param_0){
    
        .reg .b32   %r<15>;
        .reg .b64   %rd<5>;
        .reg .pred  %p<2>;
        ld.param.u64    %rd1, [Div_param_0];
        mov.u32         %r3, 3;
        mov.u32         %r4, 4;
        st.global.u32   [%rd4 + 12], 0;
        mov.u32         %r15, -1000000;
        
    BB0_1:
        add.u32         %r4, %r4, 1; 
        add.u32         %r3, %r3, 1; 
    
        div.u32         %r9,  %r4,  %r3;
        div.u32         %r10, %r3,  %r9;
        div.u32         %r11, %r9,  %r10;
        div.u32         %r12, %r10, %r11;
        div.u32         %r13, %r11, %r12;
    
        ld.global.u32   %r9, [%rd4 + 12];
        add.u32         %r10, %r9,  %r13; 
        st.global.u32   [%rd4 + 12], %r10;
        add.u32         %r15, %r15, 1;
        setp.ne.u32     %p1,  %r15, 0;
        @%p1 bra        BB0_1;
        st.global.u32   [%rd4], %r15;
        ret;
 }
\end{lstlisting}
\caption{Unsigned Div Instruction Microbenchmark written in PTX}
\label{fig:PTX_Div}
\end{figure}

\begin{figure}[t!]
\centering
\subfigure[Integer Add kernel]{%
\label{fig:add_power1}%
\includegraphics[width=3in, height=1.9in]{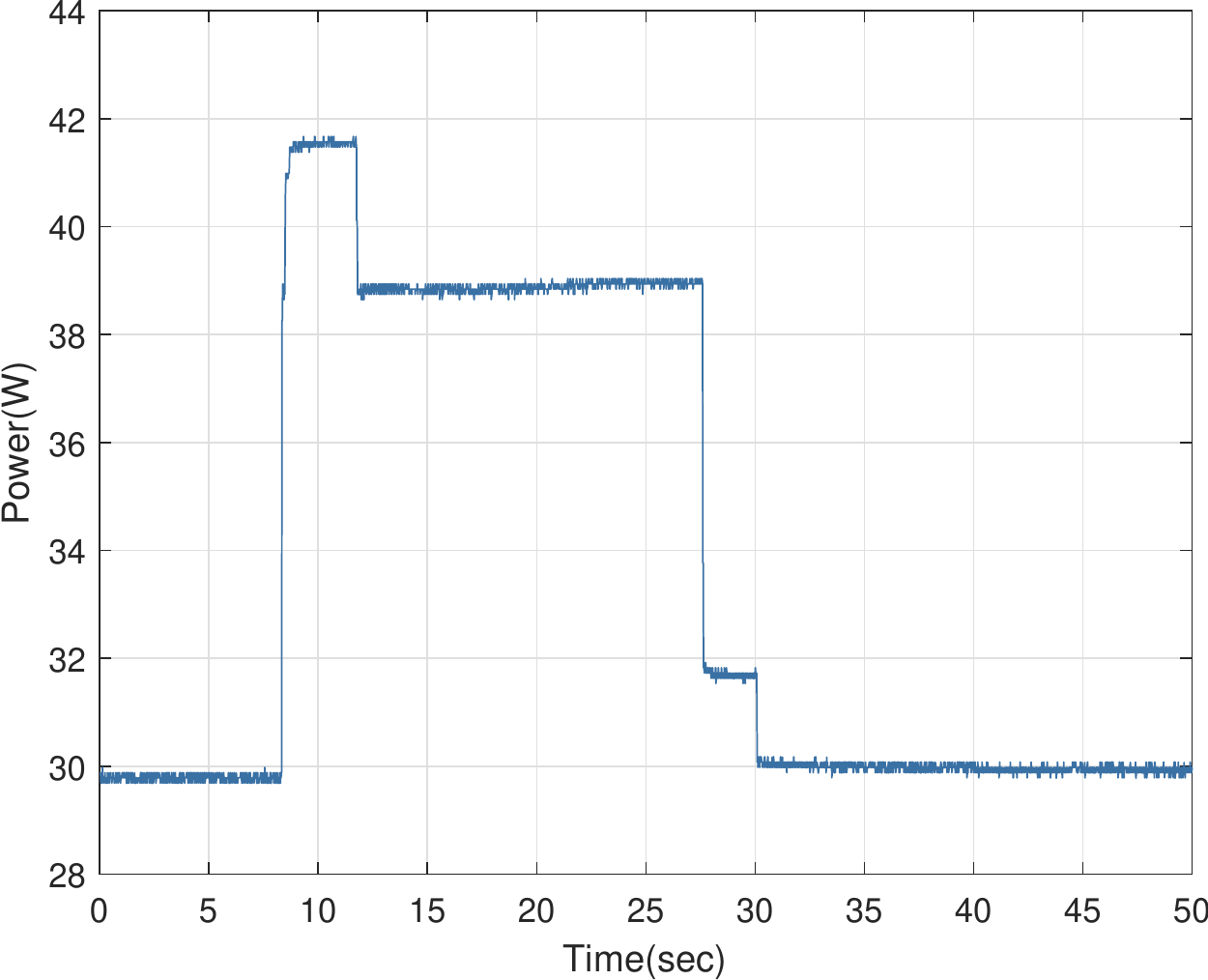}}
\qquad
\vspace{3pt}
\subfigure[Unsigned Div kernel]{%
\label{fig:div_power1}
\includegraphics[width=3in, height=1.9in]{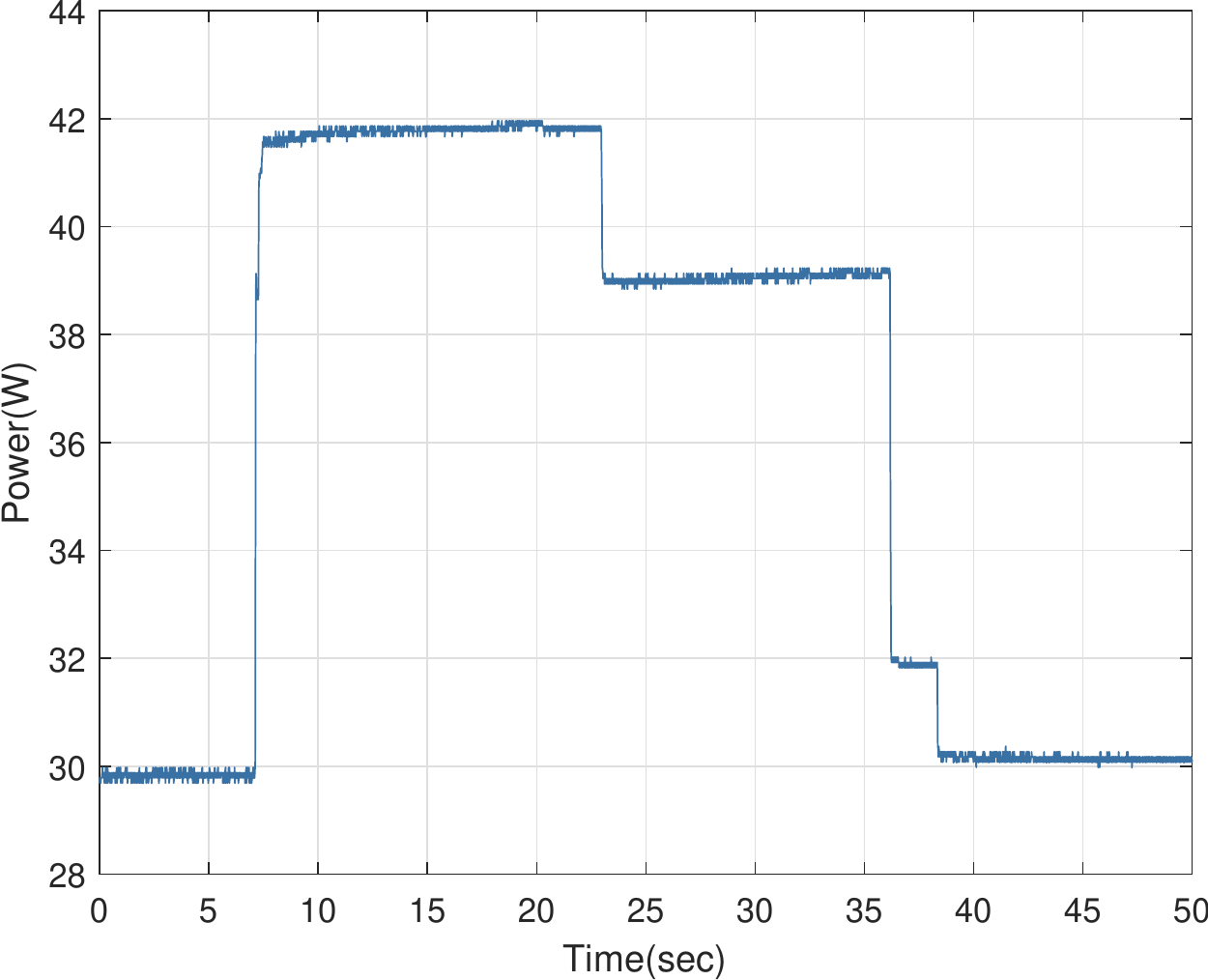}}
\caption{Add \& Div kernels Power Consumption vs time on TITAN V GPU}
\label{fig:nonAnnoted_kernels}
\end{figure}

\begin{figure*}[t!]
\centering
\subfigure[Integer Add kernel]{%
\label{fig:add_power2}%
\includegraphics[width=3.3in , height=2.1in]{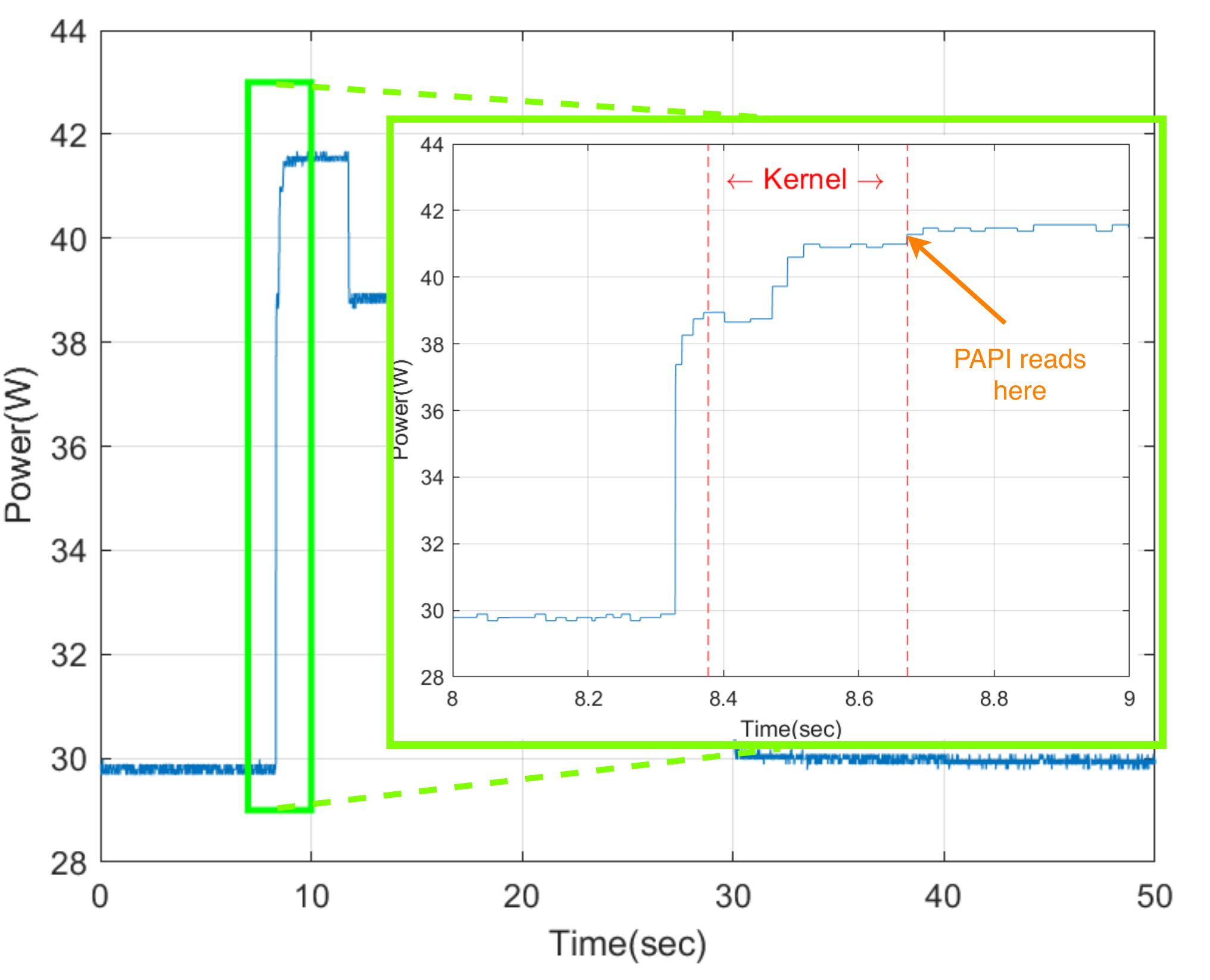}}
\qquad
\subfigure[Unsigned Div kernel]{%
\label{fig:div_power2}%
\includegraphics[width=3.5in , height=2.18in]{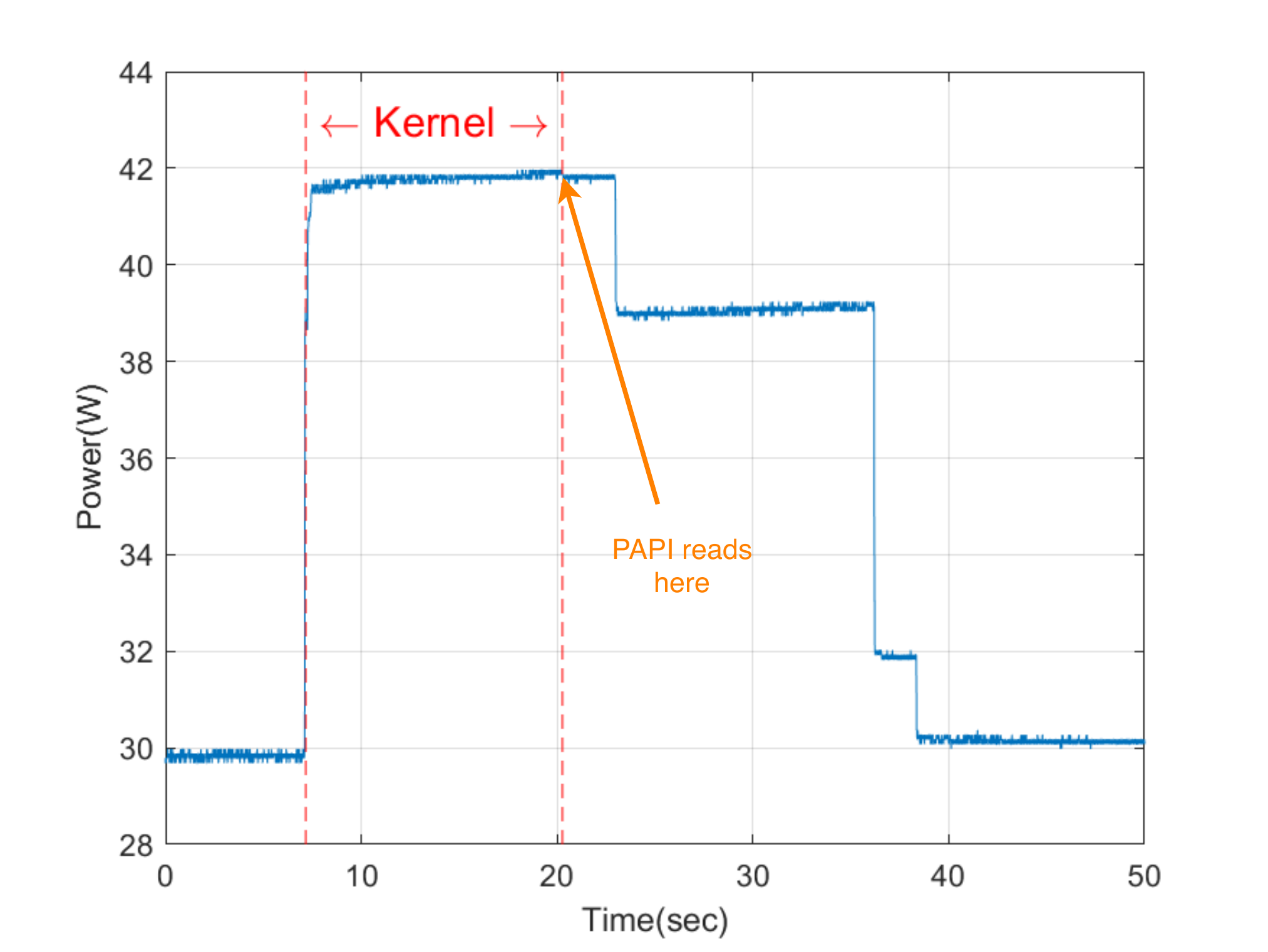}}%
\caption{Add \& Div kernels with the exact start and end of the kernel annotated}
\label{fig:annoted_kernels}
\end{figure*}

Figure~\ref{fig:PTX_Div} shows an example of the instrumented PTX kernel for the unsigned \textit{Div} instruction. In our previous work~\cite{arafaInstLatency}, we presented a similar technique to find the instruction latency. We executed the instruction only once, and red the \textit{clk} register before and after its execution. The design here is different since we need to capture the change in power usage, which would be unnoticeable if we execute the instruction only once. The key idea here is unrolling a loop and execute the same instruction millions of times and record the power then divide by the number of instructions to get the power consumption of the single instruction. The kernel in Figure~\ref{fig:PTX_Div} shows an example of the micro-benchmark of the unsigned \textit{div} instruction. We begin by initializing the used registers, lines [3--5]. Since PTX is a virtual-assembly and gets translated to the SASS, there is no limit on the number of registers to use. Still, in the real SASS assembly, the number of registers is limited and will vary from one generation/architecture to another. When the limit exceeds, register variables will be spilled to memory, causing changes in performance. Line [10] sets the loop count to 1M iterations. The loop body, lines [13--27], is composed of 5 back-to-back unsigned \textit{div} instructions with dependencies, to make sure that the compiler does not optimize any of them. We do a load-add-store operation on the output of the \nth{5} \textit{div} operation and begin the loop with new values each time to force the compiler to execute the instructions. Otherwise, the compiler would run the loop only the first time and squeeze the remaining iterations. We follow the same approach for all the instructions, and the kernel is the same, the only difference is the instruction itself.

GPUs drain power as static power and dynamic power. The static power is a constant power that the GPU consumes to maintain its operation. However, dynamic power is affected by the kernel's instructions and operations. To eliminate the static power and any overhead dynamic power other than the instruction power consumption, we measure the power and compute the kernel's energy consumption twice. First,  we run the kernel as shown in figure~\ref{fig:PTX_Div}, we call that the total energy. Second, while commenting out the back-to-back instructions (lines [16- 20]), we call that the overhead energy. We then use Eq.~\ref{eq:energy_inst} to calculate the energy of the instruction. This way, only the real energy of the instruction is calculated.

\begin{equation}\label{eq:energy_inst}
E_{instruction} = \dfrac{E_{total} - E_{overhead}}{\textnormal{\textit{\# of instructions}}} 
\end{equation}

\vspace{2pt}
\subsection{NVCC Compiler Optimization}

The kernel is compiled with (\textit{--O3}) and  (\textit{--O0}) optimization flags. This way, we capture the effect of the CUDA compiler’s higher levels of optimizations on the energy consumption of each PTX instruction. To make sure that in case of (\textit{--O3}), the compiler does not optimize the instructions and squeeze them, we made sure that the output of the kernel is correct. Line 28 of Figure~\ref{fig:PTX_Div}, stores the output of the loop. We read it and validate its correctness. Furthermore, we validate the \textit{clk} register for each instruction against our previous work~\cite{arafaInstLatency}.

\vspace{5pt}
\section{Software Measurement}
\label{sec:software}

NVIDIA provides an API named NVIDIA Management Library (NVML)~\cite{NVML}, which offers direct access to the queries exposed via the command line utility, NVIDIA System Management Interface (\textit{nvidia-smi}). NVML allows developers to query GPU device states such as GPU utilization, clock rates, GPU temperature~\etc 
Additionally, it provides access to the board power draw by querying its instantaneous onboard sensors. The community has widely used NVML since its first release with CUDA v4.1 in 2011. NVIDIA display driver is equipped with NVML, and the SDK offers the API for its use. We use NVML to read the device power usage while running the PTX micro-benchmarks and compute the energy of each instruction. There are several techniques for collecting power usage using NVML. We found that the methods do vary. Therefore, we provide an in-depth comparison of the quality of these techniques on the energy of the individual instructions.

\subsection{Sampling Monitoring Approach (\textit{SMA})}
\label{subsec:nvml1}

The C-based API provided by NVML can query the power usage of the device and provide an instantaneous power measurement. Therefore, it can be programmed to keep reading the hardware sensor with a certain frequency. This basic approach is popular and was used in other related works~\cite{k20:2014,gpujoule:2019}. The \textit{nvmlDeviceGetPowerUsage()} function is used to retrieve the power usage reading for the device, in \textit{milliwatts}. This function is called and executed by the CPU. We configured the sampling frequency of reading the hardware sensors to its maximum, 66.7 Hz~\cite{k20:2014} (15 ms window between each call to the function).

We read the power sensor according to the sample interval in the background while the micro-benchmarks are running. Example of the output using this approach are shown in Figures~\ref{fig:add_power1}  and~\ref{fig:div_power1}. The two figures show the power consumption over time for integer \textit{Add} and unsigned integer \textit{Div} kernels for the \textit{TITAN V (Volta)} GPU. The power usage jumps shortly after the launch of the kernel and decreases in steps after the kernel finishes execution until it reaches the steady-state. This is done in 22 sec and 33 sec windows interval for \textit{Add} and \textit{Div} respectively. If we calculate the two kernels actual elapsed time, it takes only 0.28 sec and 13 sec for the \textit{Add} and the \textit{Div} kernels, respectively. That is, the GPU does something before and after the actual kernel execution. Hence, identifying the window of the kernel is hard and would affect the output as the power consumption varies through time. One solution is to take the maximum reading between the two steady states, but this would be misleading for some kernels, especially the bigger ones. Therefore, we ignore this approach from reporting owing to these issues.

\subsection{PAPI API}

Performance Application Programming Interface (PAPI)~\cite{papi_c:2010} provides an API to access the hardware performance counters found on modern processors. We can read different performance metrics through either a simple programming interface from either C or Fortran programming languages. Researchers have used PAPI as a performance and power monitoring library for different hardware and software components~\cite{papi_c:2010, papi_heterogeneous:2011, papi_power:2012, papi_bluegenee:2013, papi_xeonphi:2017}. It is also used as a middleware component in different profiling and tracing tools~\cite{ldms:2014}.

PAPI can work as a high-level wrapper for different components; for example, it uses the Intel RAPL interface~\cite{rapl} to report the power usage and energy consumption for Intel CPUs. Recently, PAPI version 5.7 added the NVML component, which supports both measuring and capping power usage on modern NVIDIA GPU architectures. 

The advantage of using PAPI is that the measurements are by default synchronized with the kernel execution. The target kernel is invoked between the \textit{papi\_start}, and the \textit{papi\_end} functions, and a single number, representing the power event we need to measure is returned. The NVML component implemented in PAPI uses the function, \textit{getPowerUsage()} which query \textit{nvmlDeviceGetPowerUsage()} function. According to the documentation, this function is called only once when the \textit{papi\_end} is called. Thus, the power returned using this method is an instantaneous power when the kernel finishes execution. Although synchronizing with the kernel solves the \textit{SMA} issues, taking the instantaneous measurement when the kernel finishes execution can provide non-accurate results especially, for large and irregular kernels as shown in Section~\ref{sec:results}. Note that PAPI provides an example that works like the \textit{SMA} approach, which we refrain from this paper.

\subsection{Multi-Threaded Synchronized Monitoring (\textit{MTSM})}
\label{subsec:nvml2}

In MTSM, we identify the exact window of the kernel execution. We modified \textit{SMA} to synchronize the kernel execution. This way, only the power readings of the kernel are recorded. Since the host CPU monitors the NVML API, we use Pthreads for synchronization where one thread calls and monitors the kernel while the other thread records the power.

Algorithm~\ref{alg:energy} shows the MTSM. We initialize a volatile atomic variable (\textit{flag}) to zero, which we use later to record the power readings according to the start and end of the target kernel. On line 6 we create a new thread (\textit{th1}) which executes a function (\textit{func1}) [line 17] in parallel. This function completes the power monitoring, depending on the atomic \textit{flag}. This uses the NVML function, \textit{nvmlDeviceGetPowerUsage()} which returns the device power in milli-watts. The readings of the power during the kernel window are recorded and saved in an array (\textit{power\_readings}), which is used later in computing the kernel energy. In lines [7--12], flip the \textit{flag} value and start computing the elapsed time and the launch kernel, which means starting the power monitoring. At the end of the kernel execution, we record the elapsed time and change the flag. We use the CUDA synchronize function to make sure that the power is recorded correctly. We do not specify any reading sampling frequency for the NVML functions. Although this would give us redundant values, it would be more accurate. With this setup, we found that the power reading frequency is nearly $2 kHz$.

Figures~\ref{fig:add_power2} and~\ref{fig:div_power2} show the corresponding kernels in Figures~\ref{fig:add_power1} and~\ref{fig:div_power1} after identifying the exact kernel execution window. The new graphs are annotated with the start and end of the kernel. We observe that the kernel does not start after the sudden rise in the power from the steady-state, rather after a couple of ms from this sudden increase in power consumption (see \textit{add} kernel in Figure~\ref{fig:add_power2} for clarity). After the kernel finishes execution, the power remains high for a small-time, and then it starts descending in steps until it reaches the steady-state again. To compute the kernel’s energy, we calculated the area under the curve for the kernel using Eq.~\ref{eq:energy_area}. We believe that this approach would provide the most accurate measurement since the power readings of only the kernel are recorded. Computing the energy as the area under the curve is more rigorous than just taking the last power reading multiplied by the time elapsed for the kernel, as is done in PAPI.

\begin{algorithm}[t!]
    \caption{MTSM Approach}\label{alg:energy}
    \begin{algorithmic}[1]
        \State {\bf volatile} $elapsed\_time$, $energy$
        \State {\bf volatile atomic} $flag \gets 0$
        \Procedure{$kernel\_energy:$}{}
        \State {\bf time\_t} $time \gets 0$
        \State $pthread\_create[th_1, func_1]$
        \State $flag \gets 1$
        \State $start\_timing(\&time)$
        \State $Kernel\_call\ll Dg, Db\gg$
        \State $end\_timing(synchronize, \&time)$
        \State $flag \gets 0$
        \State $elapsed\_time \gets time$
        \State $pthread\_join(th_1)$
        \State \Return $energy$
        \EndProcedure
        
        \Procedure{$func_1:$}{}
        \State $power\_readings \gets [\;]$
        \State $monitor\_power(\&power\_readings)$
        \State $energy \gets calculate\_energy(\&power\_readings)$
        \EndProcedure
        
        \Procedure{$monitor\_power$}{$power\_readings$}:
        \While {flag}
        \State $power\_readings \gets {read\_NVML\_power\_usage}$
        \EndWhile
        \EndProcedure
    \end{algorithmic}
\end{algorithm}

We configured \textit{MTSM} as a shared library that can be linked with the application binaries at runtime. The code is first compiled and then injected or preloaded at runtime using LD\_PRELOAD~\cite{ld_preload} environment variable to any device executable binary file with a kernel that executing on NVIDIA GPUs. The start timing and end timing are automatically triggered by intercepting the CUDA runtime API~\cite{cuda-runtime-api} calls.

\begin{equation}\label{eq:energy_area}
E~(mJ) = {\dfrac{{\textnormal{\textit{Elapsed time (sec)}}}}{\textnormal{\textit{\# of power readings (N)}}}}\times{\sum_{i=0}^{N}\textnormal{\textit{Power}}_{i}~(mW)}
\end{equation}

\vspace{5pt}
\section{Hardware Measurement}
\label{sec:hardware}

Modern GPUs have two primary sources of power. The first power source is the direct DC power ($12~V$) supply, provided through the side of the card.  While the second one is the PCI-E ($3.3~V$ and $12V$) power source, provided through the motherboard. We have designed a system to measure each power source in real-time. The hardware measurement is considered as the ground truth to verify the different software measurement techniques.

\begin{figure}[tb!]
      \centering
      \includegraphics[width=\linewidth, height=0.8\linewidth]{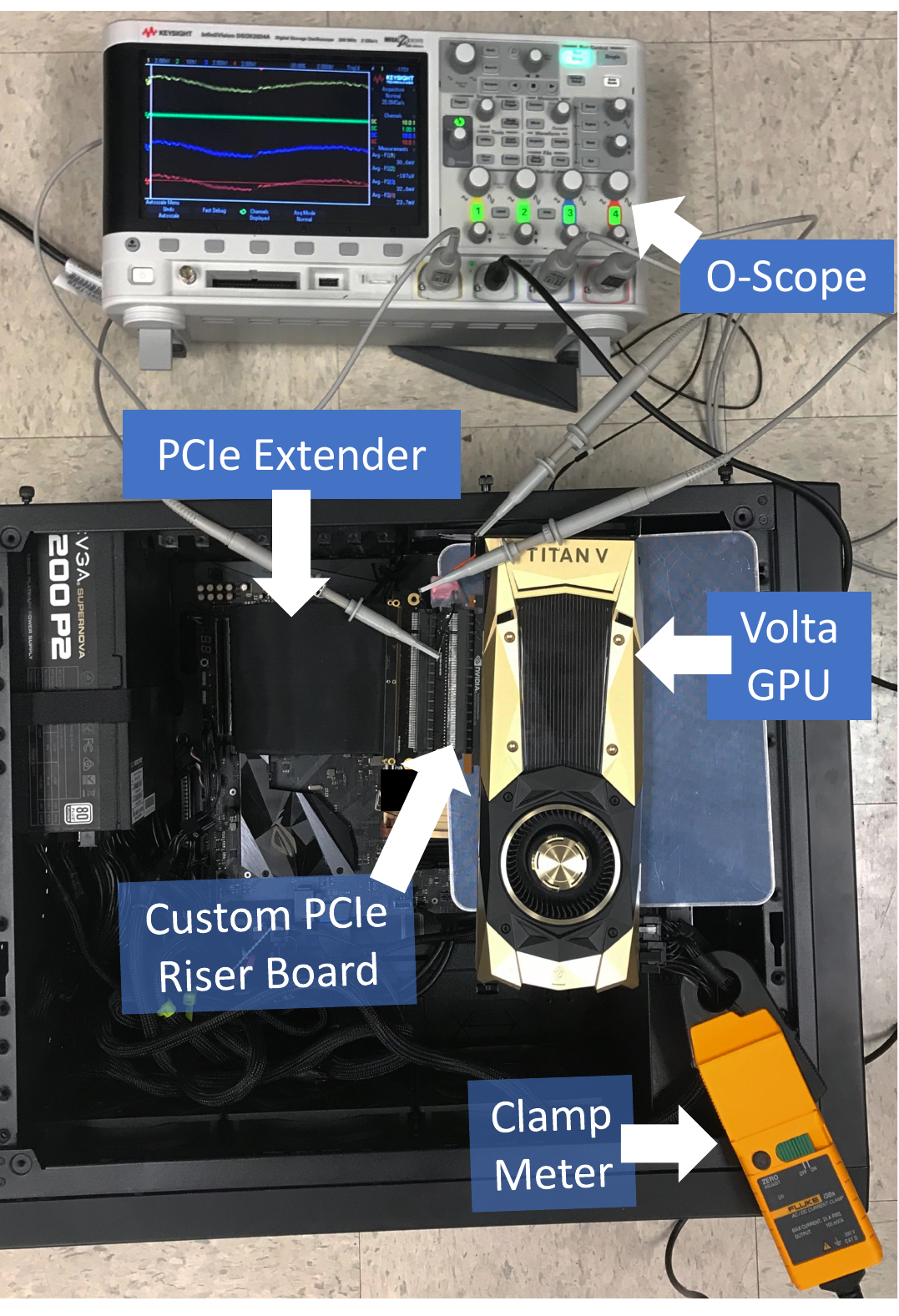}
      \caption{Hardware Measurement Setup}
      \label{fig:hardware-setup}
\end{figure}

Figure~\ref{fig:hardware-setup} shows the experimental hardware setup with all the components. To capture the total power, we measure the current and voltage for each power source simultaneously. A clamp meter and a shunt series resistor are used for the current measurement. For voltage measurement, we use a direct probe on the voltage line using an oscilloscope to acquire the signals. Equation \ref{eq:HW_power} is used to calculate the total hardware power drained by the GPU from the two different power sources.

\textbf{Direct DC Power Supply Source:} Power supply provides a $12~V$ voltage through a direct wired link. We use both a 6-pin and 8-pin PCI-E power connectors to deliver a maximum of $300~W$. Thus, the direct DC power supply source is the main contributor to the card’s power. Figure~\ref{fig:hardware-setup} shows a clamp meter measuring the current of the direct power supply connection. The voltage of the power supply is measured using an oscilloscope probe. The current and voltage are acquired using an oscilloscope, as shown in Figure~\ref{fig:hardware-setup}. Therefore, the Direct DC power supply source is calculated using simple multiplication. The third addition term in Eq.~\ref{eq:HW_power} shows the calculation of the power which is multiplying ${\textnormal{\textit{I}}_{clamp}}$ by ${\textnormal{\textit{V}}_{DPS}}$. In which, ${\textnormal{\textit{V}}_{DPS}}$ is the voltage of the direct power supply.

\begin{figure}[tb!]
      \centering
       \includegraphics[width=\linewidth, height=0.8\linewidth]{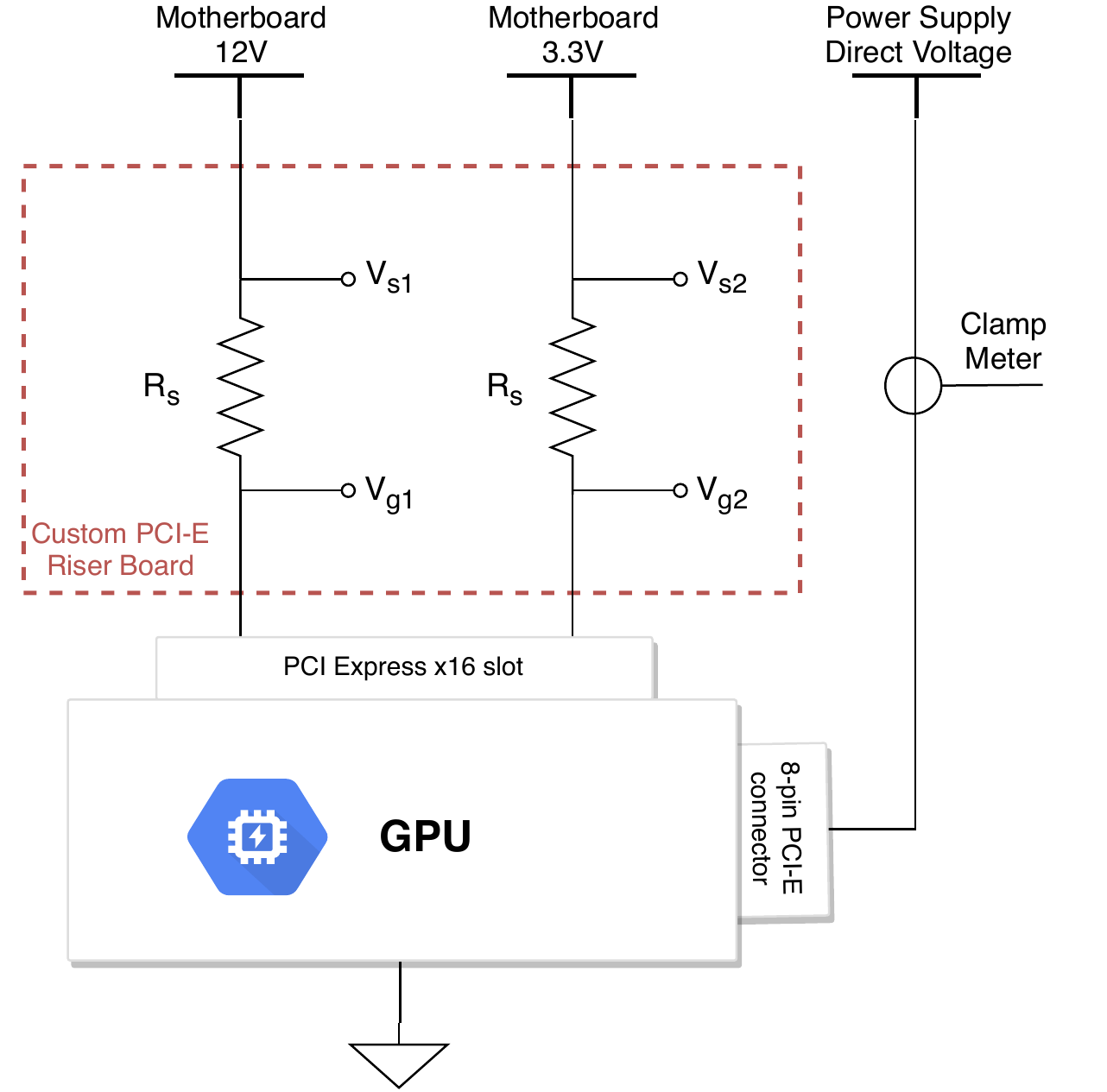}
      \caption{Circuit Diagram of Hardware Measurement}
      \label{fig:HW-sch}
\end{figure}

\textbf{PCI-E Power Source:} Graphics cards are connected to the motherboard through the PCI-E x16 slot connection. $3.3V$ and $12V$ voltages are provided through this slot. To accurately measure the power that goes through this slot, an intermediate power sensing technique should be installed between the card and the motherboard. We designed a custom made PCI-E riser board that measures the power supplied through the motherboard. Two in-series shunt resistors are used as a power sensing technique. As shown in Figure~\ref{fig:HW-sch}, each shunt resistor (${\textnormal{\textit{R}}_{S}}$) is connected in series with $3.3V$ and $12V$ separately. Using the series property, the current that flows through the ${\textnormal{\textit{R}}_{S}}$ is the same current that goes to the graphics card. Therefore, we measure the voltages ${\textnormal{\textit{V}}_{S1}}$ and ${\textnormal{\textit{V}}_{G1}}$ which are across ${\textnormal{\textit{R}}_{S}}$ using oscilloscope. We then divide it with the ${\textnormal{\textit{R}}_{S}}$ value. The voltage level is measured using the riser board. We duplicate the same calculation technique for the $3.3V$ voltage level, as shown in Eq.~\ref{eq:HW_power}.

\begin{equation}
\label{eq:HW_power}
P_{HW} = {\dfrac{{\textnormal{\textit{V}}_{s1}}-{\textnormal{\textit{V}}_{g1}}}{{\textnormal{\textit{R}}_{s}}}}\times{{\textnormal{\textit{V}}_{g1}}}+{\dfrac{{\textnormal{\textit{V}}_{s2}}-{\textnormal{\textit{V}}_{g2}}}{{\textnormal{\textit{R}}_{s}}}}\times{{\textnormal{\textit{V}}_{g2}}} + {\textnormal{\textit{I}}_{clamp}}\times{\textnormal{\textit{V}}_{DPS}} 
\end{equation}

\vspace{5pt}
\section{Results}
\label{sec:results}

We show the energy consumption of each instruction found in the latest PTX ISA, v.6.4~\cite{ptx}. We report the results of using \textit{MTSM} and \textit{PAPI} on four different NVIDIA GPUs from four different generations/architectures; \textbf{\textit{GTX TITAN X:}} GPU~\cite{gtxtitanx} from Maxwell architecture. It has 3584 cores with 151 MHz clock frequency. \textbf{\textit{GTX 1080 Ti:}} GPU~\cite{gtx1080ti} from Pascal architecture. It has 3584 cores with 1481 MHz clock frequency. \textbf{\textit{TITAN V:}} GPU~\cite{v100} from Volta architecture. It has 5120 cores with 1200 MHz clock frequency. \textbf{\textit{TITAN RTX:}} GPU~\cite{titanRTX} from Turing architecture~\cite{turing}. It has 4608 cores with 1350 MHz clock frequency.

We used CUDA NVCC compiler v.10.1~\cite{nvcc10.1} to compile and run the codes. CUDA compiler comes equipped with NVML library~\cite{NVML}. Table~\ref{table:alu_results} shows an enumeration of the energy consumption of the various ALU instructions for the different GPUs. For simplicity, we used each GPU generation to refer it. We denote the (\textit{O3}) version as \textit{Optimized} and the (\textit{O0}) version as \textit{Non-Optimized}.

The results show that overall \textit{Volta} GPUs have the lowest energy consumption per instruction among all the tested GPUs. \textit{Pascal} preceded the \textit{Volta} while \textit{Maxwell} and \textit{Turing} are power hungry devices except for some categories of the instructions.

For \textit{Half Precision} (FP16) instructions, \textit{Volta} and \textit{Turing} have much better results than \textit{Pascal}. Hence, this confirms that both architectures are suitable for approximate computing applications (\eg, deep learning, and energy-saving computing). We did not run FP16 instructions on \textit{Maxwell} as \textit{Pascal} architecture was the first GPU that offered FP16 support. The same trend can be found in \textit{Multi Precision} (MP) instructions where \textit{Volta} and \textit{Pascal} have better energy consumption compared to the two other generations. MP~\cite{multiprecisions} instructions are essential in a wide variety of algorithms in computational mathematics (\ie, number theory, random matrix problems, experimental mathematics). Also, it is used in cryptography algorithms and security. 

Overall, the energy of \textit{Non-Optimized} is always more than the \textit{Optimized}. One reason is that the number of cycles at the \textit{(O0)} optimization level are more than the \textit{(O3)} level~\cite{arafaInstLatency}. This can be because the translation from PTX instruction to native SASS instruction is not one-to-one conversion. Thus, the instruction can take more time to finish execution if it got translated to more than one instruction.

\begin{figure}[t!]
      \centering
      \includegraphics[width=\linewidth]{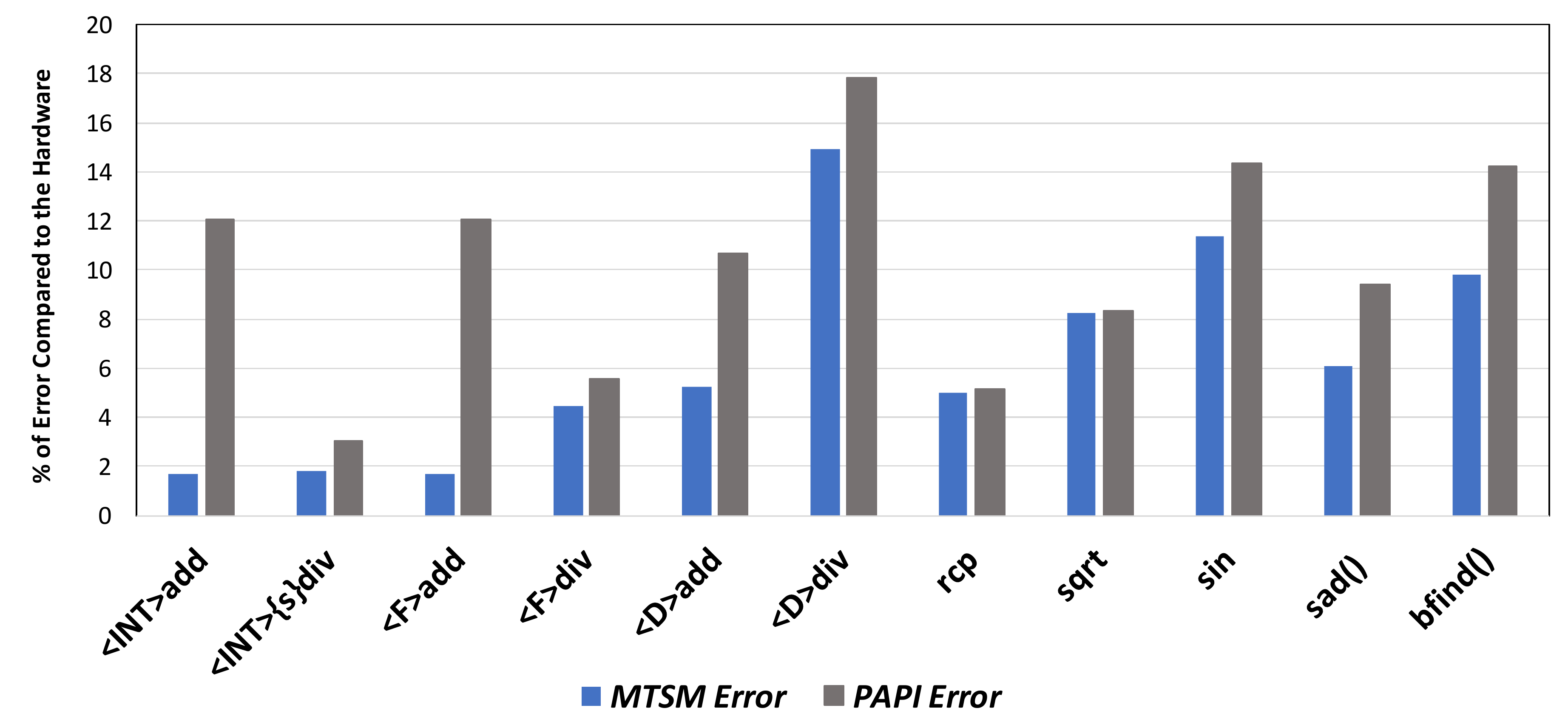}
      \caption{Instructions-level verification of \textbf{MTSM} \& \textbf{PAPI} against the \textbf{HW} on \textit{Volta TITAN V} GPU. $<$Int$>$, $<$F$>$ \& $<$D$>$ denote \textit{Integer}, \textit{Double} and \textit{Float} instructions respectively}
      \label{fig:Hardware_bar}
\end{figure}

\textbf{PAPI vs. MTSM:} The dominant tendency of the results is that PAPI readings are always more than the \textit{MTSM}. Although the difference is not significant for small kernels, it can be up to 1~$\mu$J for bigger kernels like Floating Single and Double Precision \textit{div} instructions.

\subsection{Verification with the Hardware Measurement}

We verified the different software techniques (MTSM \& PAPI) against the hardware setup on \textit{Volta TITAN V} GPU. Compared to the ground truth hardware measurements, for all the instructions, the average Mean Absolute Percentage Error (MAPE) of \textit{MTSM} Energy is \textbf{6.39} and the mean Root Mean Square Error (RMSE) is \textbf{3.97}. In contrast, PAPI average MAPE is \textbf{10.24} and the average RMSE is \textbf{5.04}. Figure~\ref{fig:Hardware_bar} shows the error of \textit{MTSM} and PAPI relative to the hardware measurement for some of the instructions. The results prove that \textit{MTSM} is more accurate than PAPI as it is closer to what has been measured using the hardware.

\vspace{3pt}
\section{Related Work}
\label{sec:related}
Several works~\cite{Power_aware:2012,Analytical3_roofline:2013} in the literature tried to directly measure the instantaneous power usage of the GPUs using various profiling techniques. On the other hand, Researchers have proposed different techniques~\cite{Analytical2_Hong:2010,gpuwattch,GPUSimPow} to indirectly estimate and predict the total GPU's power/energy consumption. Additional details are discussed by Bridges~\etal~\cite{2016_survey_power}. 

GPU power profiling can be carried out in two different approaches, a software-oriented solution, where the internal power sensors are queried using NVML, and hardware-oriented solutions using external hardware setups.

\textbf{Software-oriented approaches:} Arunkumar~\etal~\cite{gpujoule:2019} used a direct NVML sampling motoring approach running in the background while using a special micro-benchmark to calculate basic compute/memory instructions energy consumption and feed that to their model. They run their evaluation on (\textit{Tesla K40}) Kepler GPU. They intentionally disabled all compiler optimizations and compiled their micro-benchmarks with (\textit{--O3}) flag. Burtscher~\etal~\cite{k20:2014} analyzed the power consumption measured by NVML for (\textit{Tesla K20}) GPU. Kasichayanula~\etal~\cite{Power_aware:2012} used NVML to calculate the energy consumption of some GPU units which drive their model and validate it with a Kill-A-Watt power meter. While these types of hardware power meters are cheap and straightforward to use, they do not give an accurate measurement, especially in HPC settings.

\textbf{Hardware-oriented approaches:} Zhao~\etal~\cite{POIGEM:2013} used an external power meter on an old GPU (\textit{GeForce GTX 470}) from Fermi~\cite{fermi} architecture, where they designed a micro-benchmark to compute the energy of some PTX instructions and feed that into their model. The authors of~\cite{Analytical3_roofline:2013} validate their roofline model by using PowerMon 2~\cite{powermon} and a custom PCIe inter-poser to calculate the instantaneous power of (\textit{GTX 580}) GPU.

Recently, Sen~\etal~\cite{quality:2018} assessed the quality and performance of the power profiling mechanisms using hardware and software techniques. They compared a hardware approach using PowerInsight~\cite{powerinsight} (a hardware power instrumentation product) to the software NVML approach on a developed matrix multiplication CUDA benchmark. 

In a similar spirit, we follow the same line of research. Nevertheless, we focus on the energy consumption of individual instructions while having a detailed comparison of the different software/hardware approaches.

\vspace{5pt}
\section{Conclusion \& Future Directions}
\label{sec:conc}

In this paper, we accurately measure the energy consumption of various PTX instructions that execute on NVIDIA GPUs. We also show the effects of different optimization levels of the CUDA (\textit{NVCC}) compiler on energy consumption of each instruction. We provide an in-depth comparison of various software techniques that query the onboard internal GPU sensors and verify against an in-house  custom-designed hardware power measurement. Overall, the paper provides an easy and straightforward way (\textit{Multi-Threaded Synchronized Monitoring (MTSM)}) that can be used to measure the energy consumption of any NVIDIA GPU kernel\footnote{The source code is available on our laboratory page on Github at \textcolor{blue}{\url{https://github.com/NMSU-PEARL/GPUs-Energy}}.}. Furthermore, the results give GPU architects and developers a  concrete understanding of NVIDIA GPUs' microarchitecture. This work  will help GPU modeling frameworks~\cite{PPT-GPU} to have a precise prediction of energy/power consumption of GPUs. Along with GPU/CPU memory~\cite{arafaIPCCCMemoryModel,arafaICSMemoryModel,chennupati2017analytical} and pipeline~\cite{chennupati2019scalable} models, a heterogeneous system can be accurately modeled~\cite{ppt}.

\vspace{5pt}
\section*{Appendix: Energy Consumption Results}
Table~\ref{table:alu_results} has per-instruction energy breakdown for different generations of NVIDIA GPUs.

\newcolumntype{P}[1]{>{\centering\arraybackslash}p{#1}}
\renewcommand{\arraystretch}{1.5}
\begin{table*}\scriptsize
\centering
  \caption{Energy Consumption of GPU Instructions. \{s\} \& \{u\} denote signed and unsigned instructions respectively. The first number in the results is PAPI and the second one is the MTSM. All the numbers are in ($\mu$J).}
  \noindent\makebox[\textwidth]{
  \begin{tabular}{|P{1.5cm} || c | c | c | c || c | c | c | c |}
  \hline
   \multirow{4}{*}{\textbf{\textit{Instruction}}} &\multicolumn{4}{c||}{\textbf{\textit{Optimized}}} & \multicolumn{4}{c|}{\textbf{\textit{Non-Optimized}}} \\
  \cline{2-9}
  &\textbf{Maxwell} & \textbf{Pascal} & \textbf{Volta} &\textbf{Turing} & \textbf{Maxwell} & \textbf{Pascal} & \textbf{Volta} &\textbf{Turing}\\
  \cline{2-9}
  &\multicolumn{8}{c|}{\textbf{\textit{(1) Integer Arithmetic Instructions}}} \\
  \hline
   add / sub / min / max    & 0.0942 , 0.0461   & 0.0277 , 0.0200     &  0.0064 , 0.0012   & 0.0293 , 0.0281   & 1.2453 , 1.0264  &   0.6509 , 0.6203   & 0.2531 , 0.2384    & 0.8340 , 0.7905\\
  \hline
   mul / mad                & 3.0239 , 2.7309  & 0.2853 , 0.1727   & 0.0092 , 0.0014  & 0.0434 , 0.0233   & 4.5826 , 4.2959 & 3.6194 , 3.5986    & 0.5228 , 0.4912     & 0.9969 , 0.9675      \\
  \hline
  \{s\} div                 & 10.5921 , 6.7819 & 5.0270 , 4.9889 &  4.2489 , 4.0660 & 7.2119 , 6.5499 & 64.7649 , 64.6306  & 44.5100 , 44.5609  & 27.4008 , 27.0584  & 48.7700 , 48.4940  \\
  \hline
  \{s\} rem                 & 7.8512 , 6.6833 &  5.0539 , 4.9687   & 4.2100 , 4.0138 & 7.3197 , 6.7982 & 61.1036 , 61.3000   & 42.4800 , 42.0521  & 25.4413 , 25.1175    & 48.3881 , 47.9075  \\
  \hline
  abs                       & 0.800 , 0.747 &  0.2927 , 0.2349 & 0.0647 , 0.0621  & 0.3000 , 0.2710   & 2.1611 , 1.8841  & 1.2170 , 1.2448   & 1.4263 , 1.4013    &   2.3084 , 2.3880     \\
  \hline
  \{u\} div                 & 7.44783 , 6.2899  &   4.7398 , 4.5889 & 3.9254 , 3.8706 &  6.6068 , 6.038 & 52.5558 , 52.3220 & 36.2400 , 36.0963  & 20.4411 , 20.2517    & 35.8200 , 35.6736   \\
  \hline
  \{u\} rem                 & 7.5357  , 6.4006 & 4.8380 , 4.7603   & 3.9587 , 3.9471 & 6.8026 , 6.3093 & 50.6491 , 50.4818 & 35.0700 , 34.8370   & 19.6811 , 19.2906     & 35.1347 , 35.0062  \\
  \hline
  &\multicolumn{8}{c|}{\textbf{\textit{(2) Logic and Shift Instructions}}} \\
 \hline
   and / or / not / xor    & 0.0942 , 0.0461   & 0.0277, 0.0200     &  0.0064 , 0.0012   & 0.0293 , 0.0281   & 1.2453 , 1.0264  &   0.6509 , 0.6203   & 0.2531 , 0.2384    & 0.8340 , 0.7905\\
  \hline
   cnot     & 0.3362 , 0.0343 & 0.3227  , 0.2423   & 0.0071 , 0.0077  & 0.2840 , 0.1011 & 2.0562 , 1.7680  & 1.8762  , 1.8498   & 2.3421 , 2.3174   & 3.9990 , 3.9181  \\
  \hline
   shl / shr    & 0.0942 , 0.0461   & 0.0277 , 0.0200     &  0.0064 , 0.0012   & 0.0293 , 0.0281   & 1.2453 , 1.0264  &   0.6509 , 0.6203   & 0.2531 , 0.2384    & 0.8340 , 0.7905\\
  \hline
  &\multicolumn{8}{c|}{\textit{\textbf{(3) Floating Single Precision Instructions}}} \\
 \hline
   add / sub / min / max  & 0.0942 , 0.0461   & 0.0277 , 0.0200     &  0.0064 , 0.0012   & 0.0293 , 0.0281   & 1.2453 , 1.0264  &   0.6509 , 0.6203   & 0.2531 , 0.2384    & 0.8340 , 0.7905\\
  \hline
   mul / mad / fma   & 3.0239 , 2.7309  & 0.2778 , 0.2008     &  0.0021 , 0.0014   & 0.2933 , 0.2811   & 4.5826 , 4.2959  &   0.6509 , 0.6203   &  0.4874 , 0.4797    & 0.8340 , 0.7905\\
  \hline
   div                      & 10.6203 , 9.4351  & 6.9934 , 6.8707  & 5.1096 , 5.0355 & 8.6425 , 7.9232 & 57.2252 ,  56.6529  & 50.4741 , 49.9350 & 34.1050 , 33.3816  &  58.8700 , 58.6767\\
  \hline
  &\multicolumn{8}{c|}{\textit{\textbf{(4) Double Precision Instructions}}} \\
 \hline
   add / sub / min / max    &  2.3058 , 2.0061   & 1.8610 , 1.8606  & 0.3608 , 0.3567   & 2.6810 , 2.6176  & 3.3017 , 2.5143  & 2.0070 , 2.0586  & 0.5158 , 0.5114  & 4.6315 , 4.1623 \\
  \hline
   div                      & 30.0634 , 28.8160   & 19.6393 , 19.2843  & 3.7249 , 3.6828 & 25.7757 , 23.6016 & 101.3056 , 100.7807   &  50.3810 , 50.0800  & 31.0212  , 30.4121    & 67.4127 , 67.4025   \\
  \hline
  &\multicolumn{8}{c|}{\textbf{\textit{(5) Half Precision Instructions}}} \\
 \hline
   add / sub / mul   & NA  & 2.9601 , 2.8788  & 0.0924 , 0.0624  & 0.3740 , 0.1220 & NA & 3.5727 , 3.4259 & 0.5027 , 0.4656  & 0.9972 , 0.9631   \\
  \hline
  &\multicolumn{8}{c|}{\textbf{\textit{(6) Multi Precision Instructions}}} \\
 \hline
   add.cc / addc / sub.cc    & 0.3922 , 0.0791  & 0.3152 , 0.1492   & 0.0669 , 0.0535  & 0.1293 , 0.1065    & 1.2502 , 1.0270  & 0.6685 , 0.6317   & 0.5187 , 0.4938   & 0.9979 , 0.9680    \\
  \hline
   subc                      & 0.6934 , 0.3593 & 0.3655 , 0.3499  & 0.1006 , 0.089  & 0.4339 , 0.1677  & 2.1672 , 1.8927  &  1.2646 , 0.9002  & 0.9889, 0.952467   & 1.9107 , 1.8704   \\
  \hline
   mad.cc / madc               & 1.1575 , 0.7697 & 0.7981 , 0.6768   & 0.0730  , 0.0631   & 0.1621 , 0.1357  & 4.7049 , 4.4307 & 3.7018 , 3.6865   & 0.5165  , 0.5043  & 0.9979 , 0.9657   \\
  \hline
   &\multicolumn{8}{c|}{\textbf{\textit{(7) Special Mathematical Instructions}}} \\
 \hline
  rcp           & 6.4492 , 5.3416 & 4.1609 , 4.0320   & 2.4265 , 2.4270 & 4.3064 , 3.9514 & 18.6762 , 18.2830  & 13.1662 , 13.1460 & 10.2930 , 10.0538  & 19.3208 , 19.2601  \\
  \hline
  sqrt         & 6.3630 , 5.3923  & 4.1114 , 4.0068   & 2.4349 , 2.4219 & 4.3816 , 4.0533 & 19.0402 , 18.6694 & 13.4900 , 13.4185 & 10.5023 , 10.2700  & 19.7800 , 19.6984    \\
  \hline
  approx.sqrt & 0.8527 , 0.4961  & 0.3598 , 0.2345   & 1.2311 , 1.2076 & 2.1648 , 2.0121 & 15.9024 , 15.5452 & 10.7200 , 10.6812  & 8.3867 , 8.2517   & 15.4991 , 15.4438  \\
  \hline
  rsqrt & 0.5174 , 0.303 & 0.2573 , 0.1163   & 1.2488 , 1.2432  & 2.2491 , 2.0898 & 15.0459 , 14.6802 & 10.6800 , 10.6920 & 8.3784 , 8.2320  & 15.8300 , 15.7700  \\
  \hline
  sin / cos  & 0.3410 , 0.1507 & 0.1345 , 0.2742   & 0.5887 , 0.5867  & 1.0070, 0.9065 & 1.2927 , 0.8940 & 1.1390 , 0.8650  & 1.0046 , 0.9788    & 1.9340 , 1.8835  \\
  \hline
  lg2  & 0.5075 , 0.3098  & 0.3618 , 0.2371   & 1.2357 , 1.2287 & 2.1451 , 2.1634 & 14.6789 , 15.0598 & 10.7646 , 10.6786  & 8.4058 , 8.2500    & 15.6505 , 15.6127   \\
  \hline
  ex2  & 0.5147 , 0.3094 & 0.2383 , 0.3372 & 0.4798 , 0.4709  & 1.0188 , 0.6971 & 14.0001 , 13.6377 & 9.9840 , 9.9685  & 7.3144  , 7.2030  & 13.5252 , 13.4070  \\
  \hline
  copysign  & 0.2099 , 0.1700 & 0.2989 , 0.2339  & 0.0910, 0.0880  & 0.1627 , 0.1379  & 3.8932 , 3.5953 & 3.1134 , 3.1020   &  2.3692 , 2.3490    & 4.0546 , 3.9487    \\
  \hline
  &\multicolumn{8}{c|}{\textbf{\textit{(8) Integer Intrinsic Instructions}}} \\
 \hline
  mul24() / mad24()    & 0.3915 , 0.2939  &  0.2853 , 0.2727   & 0.2263 , 0.2119      & 0.3713 , 0.3415 & 6.7332 , 6.4263 & 4.8636 , 4.8636    & 2.3732  , 2.3249   & 4.6364 , 4.5942  \\
  \hline
  sad()       & 0.0316 , 0.015    & 0.2523 , 0.1243      & 0.0075 , 0.0038    & 0.2428 , 0.0422   & 1.2495, 1.0277 & 0.6371 , 0.6646    & 0.5158  , 0.5029    & 1.0100 , 0.9757\\
  \hline
  popc()    & 0.074 , 0.057   & 0.1347 , 0.2674   & 0.3968 , 0.3990   & 0.0815 , 0.0603  & 2.0281 , 1.7728  & 1.89123 , 1.9133  & 1.4984 , 1.4601   & 2.8428 , 2.7949   \\
  \hline
  clz()                & 0.0729 , 0.0479 & 0.2644 , 0.3339   & 0.5683 , 0.5657   & 0.3124 , 0.2817   & 2.0755, 1.7944   & 1.1670 , 1.2034    & 0.9145  , 0.8961     & 1.1554 , 1.4956   \\
  \hline
  bfind()              & 0.0488 , 0.0374    & 0.2326 , 0.3081   & 0.2915 , 0.2902    & 0.0304 , 0.0052  & 1.1997 , 0.9821  & 0.5912 , 0.6185    & 0.4688  , 0.4582    & 0.8010 , 0.7546   \\
  \hline
  \end{tabular}}
  \label{table:alu_results}
\end{table*}

\clearpage


\end{document}